\documentclass[aps,pra,reprint,showpacs,amsmath,amssymb]{revtex4-1}
\usepackage{}

\usepackage{graphicx}
\usepackage{bm}
\usepackage{amssymb}
\usepackage{amsmath}
\usepackage{amsthm}
\usepackage{amsfonts}
\usepackage{mathrsfs}
\usepackage{hyperref}

\begin{document}

\title{Anomalous Amplification of a Homodyne signal via Almost-Balanced Weak Values}

\author{Wei-Tao Liu$^{1,2,3}$}\email{wtliu@nudt.edu.cn}
\author{Juli\'{a}n Mart\'{i}nez-Rinc\'{o}n$^1$}
\author{Gerardo I. Viza$^1$}
\author{John C. Howell$^{1,4,5}$}

\affiliation{1 Department of Physics and Astronomy \& Center for Coherence and Quantum Optics, University of Rochester, Rochester, New York 14627, USA\\
2 College of Science, National University of Defense Technology, Changsha, 410073, China\\
3 Interdisciplinary Center of Quantum Information, National University of Defense Technology, Changsha, 410073, China\\
4 Institute of Optics, University of Rochester, Rochester, New York 14627, USA\\
5 Institute for Quantum Studies, Chapman University, Orange, California 92866, USA}




\begin{abstract}
The technique of almost-balanced weak values amplification (ABWV) was recently proposed [Phys. Rev. Lett. 116: 100803 (2016)]. We demonstrate this technique using a modified Sagnac interferometer, where the counter-propagating beams are spatially separated. The separation between the two beams provides additional amplification, with respect to using colinear beams in a Sagnac interferometer. As a demonstration of the technique, we perform measurements of the angular velocity in one of the mirrors of the interferometer. Within the same setup, the weak-value amplification technique is also performed for comparison. Much higher amplification factors can be obtained using the almost-balanced weak values technique, with the best one achieved in our experiments being as high as $1.2\times10^7$. In addition, the amplification factor monotonically increases with decreasing post-selection phase for the ABWV case in our experiments, which is not the case for weak-value amplification at small post-selection phases.
\end{abstract}





\maketitle 
Weak measurements, introduced by Aharonov, Albert and Vaidman \cite{prl601351}, provide a low-disturbance method to extract information of a physical system by using an auxiliary pointer. The protocol is a valuable metrological technique for parameter estimation due to the anomalous amplification related to the weak value, under properly chosen pre- and post-selection. The sensitivity and measurement precision limited by technical noise can be effectively improved, since such amplification helps by increasing the signal while decreasing or retaining the technical-noise floor \cite{rmp86307, prx4011031,G}. 
Based on weak-value amplification (WVA), ultrasensitive measurements for optical beam deflections\cite{science319787,oe1916508,pra85043809,prl102173601,ol361698,ol361479}, phase shifts \cite{prl107133603,prl111033604,14055710}, frequency shifts \cite{pra82063822}, velocities \cite{ol382949}, temporal shifts \cite{prl105010405,prl110083605}, angular shifts \cite{prl112200401}, and temperature changes \cite{ol374991} have been experimentally achieved.

The anomalous amplification of WVA in the pointer's readout is accompanied by discarding data due to pre- and post-selection in the system. At the same time, prior information about the input states is necessary to calculate the post-selection phase and the weak value. Towards these issues, Mart\'{i}nez-Rinc\'{o}n et al. proposed a novel technique dubbed almost-balanced weak values (ABWV) amplification \cite{julian}, in which the intensities at the two outputs of an interferometer are set almost equal and the difference between them reveals anomalous amplification. This technique relies on balancing two weak values instead of using one large anomalous weak value as is the case in WVA. All the photons are measured, so that no data is discarded. What's more, the differencing and sum of two outputs can be used for calculating the equivalent post-selection phase, therefore prior information about the input state of the pointer is no longer required. The background noise is also removed by differencing. Mart\'{i}nez-Rinc\'{o}n et al demostrated precision measurements for small rotation velocities in the polarization of a laser pulse. In this letter, we use ABWV amplification in a modified Sagnac interferometer to measure a constant angular velocity of one of the mirrors in the interferometer. Furthermore, we experimentally compare the performance of WVA and ABWV techniques within the same experimental setup.

\begin{figure}[hbtp]
  \centering
  \includegraphics[width=\columnwidth]{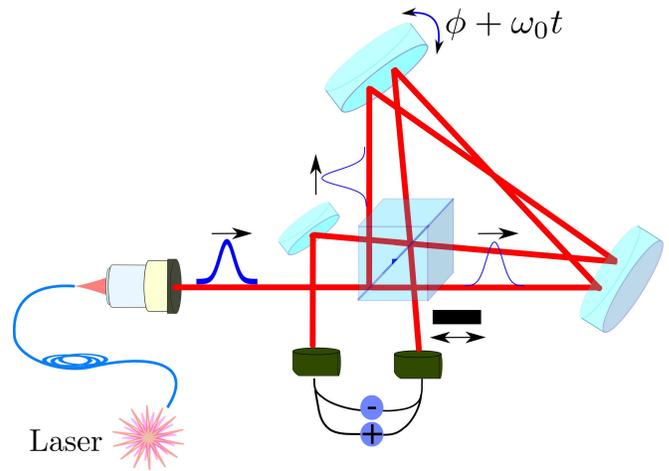}
  \caption{Illustration of the experimental setup. The interferometer consists of one beam-splitter and two mirrors, with one of them rotating with angular velocity $\omega_0=156$nrad/s. The counter-propagating beams are spatially separated, introducing a phase shift due to the tilt of the rotating mirror. Both ports are tracked, and by blocking or unblocking one of them, measurements using WVA or ABWV technique can be performed.}\label{setup}
\end{figure}

For ABWV amplification, both outputs of the interferometer are utilized. To achieve this, a modified Sagnac interferometer is employed, where the counter-propagating beams in the interferometer are spatially separated. Therefore the outputs are spatially separated from the input beam. The interferometer is set up with two mirrors and one beam-splitter, as shown in Fig.\ref{setup}. In this case, two beams are crossing each other in the long arm of the interferometer. This configuration offers tilt sensitivity due to phase difference imparted on the beams. With one mirror rotating at a small angular velocity to be measured, the weak interaction between the system and the pointer is introduced as the phase shift caused by tilting of the rotating mirror. Our definition of system and pointer here is the which-path degree of freedom in the interferometer and the time coordinate of the laser pulse, respectively. In this configuration of the interferometer, both beams will be deflected into the same direction at the outputs, while the phase difference between them is shifted according to the amplitude of the tilt. By setting the interferometer at destructive interference in one of the outputs and detecting such dark port, WVA is performed. When the intensities of the two outputs are set almost balanced, both outputs are detected and ABWV amplification can be performed.

A 795-nm continuous-wave laser beam (Vescent Photonics distributed Bragg reflector laser diode D2-100-DBR) is sent through an Acousto-Optic Modulator (AOM). The AOM is used to modulate a Gaussian profile in the field's intensity, i.e. $I\propto\exp\left(-t^2/2\tau^2\right)$, which is coupled to a single mode patch cable to clean the spatial mode. The non-Fourier band-limited pulses with repetition rate $f_r=1$Hz are sent to the interferometer composed of one 2-inch 50/50 beam-splitter and two 2-inch mirrors.

A piezoelectric actuator (Thorlabs PE4) is adapted to one of the mirrors to induce the time-dependent tilt, $(\phi+\omega_0t)$, in the horizontal direction. The origin of $\phi=0$ for WVA is determined by the mirror's position where destructive interference occurs, and where the two outputs are entirely balanced for the case of ABWV. The angle $\phi$ controls the peak output intensity on both ports of the interferometer, and the constant angular velocity $\omega_0$ is the parameter to be estimated during the experiments. The tilt induces a phase difference of $2k_0L(\phi+\omega_0t)$ between both arms of the interferometer, where $k_0=2\pi/\lambda$ and $L=L_1\cos\theta_1-L_2\cos\theta_2$, with labels 1 and 2 referring to the counter-propagating beams. $L_i$ is the distance from the pivot in the mirror mount to the place where the $i^{th}$ beam hits the mirror, and $\theta_i$ is the incident angle of the $i^{th}$ beam with respect to the mirror's normal. For example, if the interferometer was a perfect $45^\circ,45^\circ,90^\circ$ triangle (which it was not), these angles would be $\theta_1\approx\theta_2\approx22.5^o$, and $L\approx l\cos(22.5^\circ)$, where $l=L_1-L_2$ 
 is the distance between both beams on the surface of the piezo-driven mirror. The effective value of $L=5.64\pm0.03$ mm was independently measured by applying a 300 mV peak-to-peak 1-Hz signal on the piezo actuator, and using the known piezo response of $\alpha=3.12$ $\mu$rad/V. During the experiment, the angle $\phi$ was set such that $\omega_0\tau\ll\phi$ and $2k_0L\phi\ll1$ to satisfy the weak-value approximation. A 60$\%$ duty-cycle triangle ramp with peak-to-peak voltage $V_{pp}$ and frequency $f_r$ was applied to the piezo actuator, so the angular velocity of the mirror during the positive ramp takes the form of $\omega_0=5\alpha V_{pp}f_r/3$. 

The output beams of both ports of the interferometer were directed to two of the four detectors in a quadrant cell photoreceiver (Newport 2921), which outputs signals equivalent to the sum and difference intensities. By blocking or unblocking one of the two optical ports and controlling the angle $\phi$, the system resembled either the WVA or the ABWV technique respectively. The difference and sum electrical signals were sent to two, connected-in-series, low-noise voltage preamplifiers (Standard Research Systems SR560), before being recorded using an oscilloscope and a computer. These preamplifiers were set as 12 dB/oct rolloff low-pass filters at 30 Hz. The experiments were performed at 1 Hz, with no frequency filtering close to the working frequency.

According to the theory of ABWV amplification \cite{julian}, the sum and difference signals are given by

\begin{equation}\label{sum}
I_+(t)=I_0\,e^{-t^2/2\tau^2},
\end{equation}
\begin{equation}\label{abwva}
I_-(t)=I_0\,\sin(2k_0L\phi)\,e^{-(t-\omega_0\tau^2/\phi)^2/2\tau^2}.
\end{equation}
That is, the differencing signal shows a time shift of $\omega_0\tau^2/\phi$ with respect to the sum signal, including an amplification factor of $1/\phi$.

For the case of WVA, the intensity at the dark port is 
\begin{equation}\label{wv}
I_{WVA}=I_0\sin^2(k_0L\phi)\,e^{-(t-2\omega_0\tau^2/\phi)^2/2\tau^2},
\end{equation} 
where there is a time shift of $2\omega_0\tau^2/\phi$ compared to the input signal. Here $\sin^2(k_0L\phi)$ is the probability of postselection, i.e. only $N\sin^2(k_0L\phi)$ photons are detected in the dark port for WVA, with $N$ being the number of input photons to the interferometer.

\begin{figure}
  \centering
  \includegraphics[width=\columnwidth]{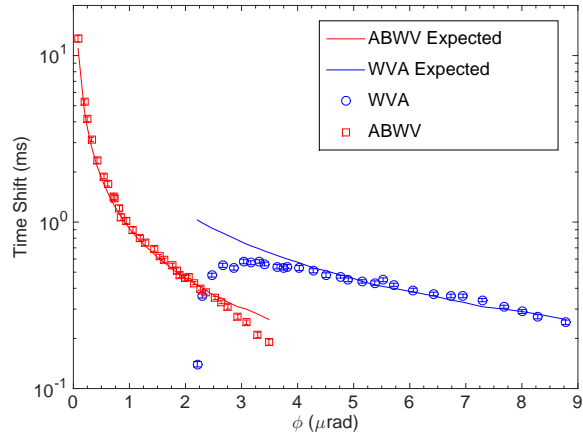}
  \caption{Time shifts for WVA (blue circles) and ABWV (red squares) techniques. The values for $\phi$ and time shifts are all obtained from experimental data by Gaussian fitting. Each point shows the mean and the standard deviation for statistics over 60 pulses. The solid lines show the expected time shift as function of $\phi$.}\label{timeshift}
\end{figure}

The outputs of the preamplifiers are then measured on an oscilloscope for data collection. Data from 60-second sets (each containing 60 pulses) from the oscilloscope are recorded on the computer for postprocessing. Each collection window consists of three million points, giving a time resolution of 20 $\mu$s. Each collected pulse is numerically fitted to a Gaussian distribution, such that peak values, characteristic lengths of pulses $\tau$, and time shifts are obtained. The results of time shifts are shown in Fig.\ref{timeshift}, where each point shows the result of statistics over a set of 60 pulses. 

\begin{figure}
  \centering
  \includegraphics[width=\columnwidth]{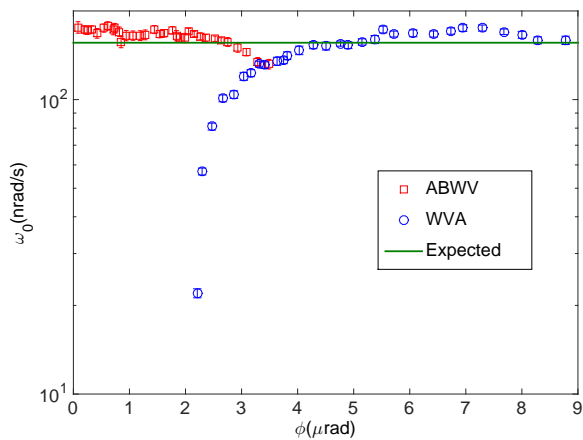}
  \caption{Angular velocity $\omega_0$ retrieved from experimental results for WVA (blue circles) and ABWV (red squares) techniques. The solid lines show the expected value of $\omega_0$, calculated according to the response of the piezo actuated mirror mount and the parameters of applied ramp.}\label{omega}
\end{figure}

For ABWV, the sum and difference signals are both analyzed. By fitting the measured intensities to Eqs.(\ref{sum}) and (\ref{abwva}), the phase $2k_0L\phi$ and the time shift can be determined. The measurement results for $\omega_0$ can then be extracted. For WVA, prior information about the input is necessary to obtain the post-selection phase and the time shift, which can be obtained by setting the phase to constructive interference and detecting the bright port separately. Results from both techniques are shown in Fig.\ref{omega}. Similar accuracy is observed within each technique's observed well-behaved intervals. These intervals are $\phi\in[83$nrad, 2.5$\mu$rad] for ABWV and $\phi\in[4\mu$rad,9$\mu$rad] for WVA.

Comparing Eq.(\ref{abwva}) and Eq.(\ref{wv}), we notice that ABWV technique provides higher detectable intensities, since the quadratic form in WVA, $\sin^2(k_0L\phi)\sim (k_0L\phi)^2$ is replaced with a larger linear response $\sin(2k_0L\phi)\sim 2k_0L\phi$.

From the results, with ABWV we can perform measurements with smaller angles $\phi$ than with WVA by more than one order of magnitude. This allows one to achieve larger amplification factors thus larger time shifts (Fig.\ref{timeshift}).
Also, ABWV technique does not show the undesired reversed tendency of WVA for the time shift at small angles $\phi$, which also allows for larger amplification. The WVA technique breaks down at around $\phi\sim 4\mu$rad, while ABWV behaves well down to $\phi\sim$83 nrad. For WVA, the strong falling off behavior is mainly caused by imperfections of optics \cite{concatenated}. These imperfections can be eliminated by using the Homodyne-like differencing response. For the ABWV technique, the two outputs are set almost balanced, which allows to combine the advantages of both techniques: elimination of common mode noise because of Homodyne detection and amplification inversely proportional to the post-selection angle as in the weak-value amplification technique. At the same time, intensity of possible unexpected background noise induced by imperfect optics is much smaller than the intensity of each output, therefore the relative error (intensity of background noise over detected intensity) for ABWV case is much smaller than that of WVA. In this point of view, ABWV is less sensitive to imperfections of the interferometer than WVA. 

It should be noted that the largest amplification factor achieved in our experiments is $1/83nrad\sim1.2\times10^7$. As a contrast, the amplification factors are usually at the level of $10^2$ for WVA using regular Sagnac interferometer, with the best post-selection phases being around tens of mrad \cite{G,concatenated}. For weak-value experiments, the amplification factor is usually equal to the reciprocal of the post-selection phase. In our case, from Eq.(\ref{abwva}), the post-selection phase is $2k_0L\phi$. However, the amplification factor is $1/\phi$ instead of the reciprocal of the post-selection phase. A significant enhancement in the amplification factor, by a factor of $2k_0L\sim 10^5$, is induced using the modified Sagnac interferometer. The rotation angle of the mirror is transferred into phase difference between two beams, being magnified at the same time because of the spatial separation of two beams.

In conclusion, we experimentally demonstrated ABWV measurements with a modified Sagnac interferometer. The separation between two counter-propagating beams provides an additional amplification factor, which can be useful for precision measurement of parameters related to angular rotations. Precision measurements of a constant angular velocity were performed, using two techniques, WVA and ABWV, within the same experimental setup. It was observed that ABWV can achieve much larger amplification. Also, we can infer that ABWV technique is less sensitive to imperfection of interference than WVA, and the reversed tendency of WVA when the post-selection phase is smaller than a certain value does not show up in our experiments. In addition, both sum and differencing signals can be obtained from the detection results thus no prior information of the input pulses is requested, which makes the estimation free of systematic errors. For our specific experiments, we saw the maximum amplification for the ABWV technique 24 times \cite{com} larger than the best result for the WVA technique. We
believe ABWV technique will find interesting applications on
precision metrology, within and outside optical systems.

This work was supported by the Army Research
Office, Grant No.W911NF-12-1-0263, Northrop Grumman
Corporation, and by National Natural Science Foundation of China (NSFC)(11374368). W. T. Liu is also supported by the Program for New Century Excellent Talents in University.


\begin{thebibliography}{99}
\bibitem{com} The amplification factor is $1/\phi$ in case of ABWV and $2/\phi$ for WVA, respectively. The best observed $\phi$ is 83nrad for ABWV and 4$\mu$rad for WVA. Therefore the maximum amplification from ABWV is $4\mu$rad$/$(83nrad*2)$\sim24$ times larger than that of WVA.
\end{thebibliography}
\end{document}